\input harvmac

\def\p{\partial}
\def\ap{\alpha'}

\def\X{{\bf X}}
\def\Y{{\bf Y}}

\Title{EFI-99-4}{\vbox{\centerline{Ten Dimensional Black Hole and 
}
\vskip12pt
\centerline{the D0-brane Threshold Bound State}}}
\vskip20pt

\centerline{Miao Li}
\centerline{\it Enrico Fermi Institute}
\centerline{\it University of Chicago}
\centerline{\it 5640 Ellis Avenue, Chicago, IL 60637} 
\centerline{\tt mli@theory.uchicago.edu}

\bigskip

We discuss the ten dimensional black holes made of D0-branes in the
regime where the effective coupling is large, and yet the 11D geometry
is unimportant. We suggest that these black holes can be interpreted
as excitations over the threshold bound state. Thus, the entropy formula
for the former is used to predict a scaling region of the wave function
of the latter. The horizon radius and the mass gap predicted in this 
picture agree with the formulas derived from the classical geometry.

\Date{January 1999}

\nref\imsy{N. Itzhaki, J. M. Maldacena, J. Sonnenschein and 
S. Yankielowicz, hep-th/9802042.}
\nref\joe{J. Polchinski, unpublished.}
\nref\mm{V. Balasubramanian, R. Gopakumar and F. Larsen, 
hep-th/9712077; A. Jevicki and T. Yoneya, hep-th/9805069; S. Hyun,
hep-th/9802026;
S. Hyun and Y. Kiem, hep-th/9805136.}
\nref\bfks{T. Banks, W. Fischler, I. R. Klebanov and L. Susskind,
hep-th/9709091; hep-th/9711005; I. R. Klebanov and L. Susskind,
hep-th/9709108;
G. Horowitz and E. Martinec,
hep-th/9710217.}
\nref\lm{M. Li, hep-th/9710226; M. Li and E. Martinec, hep-th/9801070.}
\nref\lms{M. Li, E. Martinec and V. Sahakian, hep-th/9809061;
E. Martinec and V. Sahakian, hep-th/9810224; hep-th/9901135.}
\nref\jm{J. A. Minahan, hep-th/9811156.}
\nref\suss{L. Susskind, hep-th/9901079.}
\nref\bfss{T. Banks, W. Fischler, S. Shenker and L. Susskind,
hep-th/9610043.}
\nref\kl{D. Kabat and G. Lifschytz, hep-th/9806214.}
\nref\bfk{T. Banks, W. Fischler and I. Klebanov,
hep-th/9712236.}
\nref\stat{N. Ohta, J-G. Zhou, hep-th/9801023; 
M. L. Meana, M. A. R. Osorio, J. P. Penalba, hep-th/9803058.
J. Ambjorn,
Y. M. Makeenko and G. W. Semenoff, hep-th/9810170.}
\nref\gl{Y.-H. Gao and M. Li, hep-th/9810053.}

The 10 dimensional black hole consisting of D0-branes in the regime
where the classical 10 dimensional geometry defined by the string
metric is reliable has been hard to understand. This difficulty is
caused partly by the large effective coupling constant in the D0-brane
quantum mechanics, and partly by the fact that unlike other higher
dimensional D-branes, D0-branes are highly dynamical due to the
Heisenberg uncertainty relation.

Our basic observation in this paper is that the black hole in the
regime of interest can be regarded as a small excitation over the
large N threshold bound state. Thus, it is hopeless to make a first
principle calculation concerning the black hole without understanding
the bound state first. What we shall do here is to postulate that 
there is a scaling regime where the one-particle wave function obeys
a power law, matching the entropy formulas fixes the exponent.
As a consistency check, we show that the correct formula for the
horizon radius emerges naturally in this picture. We also estimate
the mass gap in the black hole background.

The near horizon geometry of a D0-brane black hole is given by
\imsy
\eqn\metr{\eqalign{ds^2&= -fdt^2+f^{-1}dU^2+\sqrt{\lambda}U^{-3/2}
d\Omega_8^2,\cr
e^\phi &=g_s\left({\lambda\over U^7}\right)^{3/4},}}
where 
\eqn\para{f={U^{7/2}\over\sqrt{\lambda}}\left(1-{U^7_0\over U^7}
\right),\quad \lambda =60\pi^3g_sN.}
We have set $\ap =1$.
It was argued by Polchinski that there is a duality between M
theory defined on the background geometry of the threshold bound
state and quantum mechanics of D0-branes in the large N limit
\joe. For related discussions, see \mm.

This duality reduces to that between the IIA theory and D0-brane
quantum mechanics when the curvature is smaller than the string
scale, and the dilaton is sufficiently small. The conditions
are just \imsy
\eqn\condi{ 1\ll \sqrt{\lambda}U_0^{-3/2}\ll N^{2/7}.}

The Hawking temperature is 
\eqn\hawt{T={1\over 4\pi}\p_U g_{00}(U=U_0)={7\over 4\pi}\lambda^{-1/2}
U_0^{5/2}.}
In the M theory unit, we have
\eqn\radius{r_0=U_0\ap =N^{1/5}l_p^{9/5}(\beta R)^{-2/5},}
where we ignored a numerical factor.
Define the effective coupling by $\lambda_{eff}=NR^3\beta^3/l_p^6$,
the entropy of the black hole is given by
\eqn\entr{S=N^2\lambda_{eff}^{-3/5},}
again a numerical factor is dropped.
Conditions \condi\ are rewritten in terms of the effective coupling
\eqn\condm{1\ll \lambda_{eff}\ll N^{10/7}.}
Since the effective coupling is large, it is hopeless to describe
this black hole using the perturbative quantum mechanics of D0-branes.
Also notice that there is a upper bound on the effective coupling,
this is to avoid the M theory region where the effective 11th dimension
becomes larger than the Planck length. Thus, the 10 dimensional
black hole we are aiming to describe in this paper is different
from the matrix black holes discussed in \refs{\bfks, \lm}.
This upper bound also indicates that for a given temperature, we do not
have to hold the 't Hooft coupling fixed. Indeed, the natural limit
in matrix theory is to hold the Yang-Mills coupling fixed.
It is unclear whether a different phase should occur when one
crosses the upper bound for $\lambda_{eff}$. The entropy of
the black hole is of order $N^{8/7}$ at this bound, this is still 
larger than $N$ where a localization transition occurs, for a 
systemacally discussion on the phase diagram, see \lms. We shall 
here adopt a conservative viewpoint, and obey the upper bound.

As in the background of any black hole in any dimension, there is
a mass gap. The mass gap is determined by the Hawking temperature.
A WKB solution of the dilaton equation of motion clearly indicates
that indeed this mass gap is order of the temperature \jm. We shall
not quote the precise formula here, since we will not try to calculate
the exact coefficient. Again, this behavior is hard to understand
in terms of the perturbative nonabelian quantum mechanics, which is valid
only in the high temperature regime. A
rather straightforward calculation shows that the one-loop mass gap
$m^2$ is actually negative, and
is proportional to $\lambda_{eff}T^2$, see the appendix.

Recently it has been argued that the size of the large N threshold bound
state grows with N as $N^{1/3}$ \refs{\joe,\suss}. The bound state
is a highly nonabelian object. It is impossible to separate the D0-brane
degrees of freedom from the off-diagonal elements. It is still reasonable
to talk about the density of D0-branes. The situation is similar
to the matrix membrane \bfss, where the object is also highly
nonabelian, nevertheless so long if N is sufficiently large,
it is meaningful to define D0-brane or the longitudinal momentum density
along the membrane.

We shall not repeat the argument which leads to the estimate of
the size of the bound state $L\sim N^{1/3}l_p$. For our purpose,
it is sufficient to know the order of magnitude of a few relevant
quantities. The size can be measured by the quantity 
$1/N\langle \tr X^2\rangle=L^2$, $X$ is one of the nine bosonic 
matrices. The typical frequency of an off-diagonal element is of 
order \suss
\eqn\freq{\omega =N^{1/3}Rl_p^{-2},}
thus the ``zero-point energy" of an off-diagonal state is
of this order. This huge amount of energy is supposed to be
canceled by contribution from the fermionic fluctuations.
If we assume that the black hole is a ``small excitation" over
the threshold bound state, the surplus energy must come 
from excitations of the off-diagonal elements. This is because the
energy is of order $N^2$ when $\lambda_{eff}$ is held fixed.
The typical energy carried by such elementary quanta is just $T$.
This is much smaller than $\omega$ since $\omega/T
=\lambda_{eff}^{1/3} \gg 1$.
Indeed this is a small fluctuation over the bound state
configuration as long as the bosonic degrees are concerned.
So the wave function will not be significantly changed by
the fluctuation. This is our basic observation.

It is easy to see that the condition $\lambda_{eff}\gg 1$
is also equivalent to the condition $r_0\ll L$. Physically,
one would imagine that it is sufficient to know the shape
of the wave function in this region in order to calculate 
the thermodynamic quantities of the black hole. This will turn out
to be true. For our purpose, it it enough to know the density
of D0-branes in this region, $\rho (r)\sim |\psi (r)|^2$. We shall
adopt the following scaling ansatz
\eqn\sca{\rho(r) =CNL^{-9}\left({r\over L}\right)^\alpha,}
where $\alpha$ is to be determined, $C$ is a numeric constant
which we will drop henceforth.  We used rotational
invariance in this ansatz. Obviously, $L$ must be
the only relevant scale in this problem. Beyond $L$, 
there may be another scaling region which can not be probed
by black hole. This is the perturbative region in quantum
mechanics, however.

The above ansatz is valid only when we smear over scales much
larger than the average spacing between two neighboring D0-branes.
As we have emphasized, it is difficult to define such spacing
with a highly nonabelian configuration. When we make an
observation only in one direction, say $X$, it makes sense
to define such a spacing. It is $L/N\sim N^{-2/3}l_p$, a
rather small quantity in the large N limit. The high density
in one dimension is caused by the projection. A naive
definition of the true spacing is $a\sim (L^9/N)^{1/9}
\sim N^{2/9}l_p$. This is a rather large quantity. Lacking
a better definition, we shall use this naive one here.

The frequency for a diagonal element stretched between a D0-brane
at point ${\bf X}$ and another D0-brane at point ${\bf Y}$ is
$R|{\bf X}-{\bf Y|}l_p^{-3}$, thus the Boltzmann weight is a 
function
\eqn\bolt{f(\beta R|{\bf X}-{\bf Y}|l_p^{-3}|).}
We shall not specify this function. For a bosonic degree of freedom,
$f(x)=(e^x -1)^{-1}$, and for a fermionic degree of freedom,
it is $f(x)=(e^x+1)^{-1}$. To include all possible states,
$f$ can be more complicated. (There is no chemical potential,
since individual elementary quanta associated to an off-diagonal
element are not conserved.) The contribution of these states
to the free energy is another function $g(x)$. For a boson, it
is $\ln (1-e^{-x})$ and for a fermion, it is $\ln (1+e^{-x})$.
The total free energy is
\eqn\freee{\beta F=\int \rho({\bf X})\rho({\bf Y})d^9X
d^9Yg(\beta R|{\bf X}-\Y|l_p^{-3}|).}
Rescaling $\X\rightarrow L\X$ and using the ansatz \sca\ we
obtain
\eqn\mfree{\beta F=N^2\int d^9Xd^9Y|\X|^\alpha |\Y|^\alpha
g(\lambda_{eff}^{1/3}|\X-\Y|).}
The entropy can be obtained using the standard formula
$S=\beta E-\beta F$. It is obvious that the entropy is a function
of $N^2$ and $\lambda_{eff}$ only, just like $\beta F$. This result 
justifies our observation that indeed the black hole is a
thermal excitation over the threshold bound state.

To obtain the right dependence on $\lambda_{eff}$, let us do a
further rescaling $\X\rightarrow \lambda_{eff}^{-1/3}\X$.
The free energy reads
\eqn\free{\beta F= N^2\lambda_{eff}^{-(2\alpha +18)/3}
\int d^9Xd^9Y|\X|^\alpha |\Y|^\alpha g(|\X-\Y|).}
Now the integral is a pure number. Matching onto the scaling
in \entr\ determines the exponent
\eqn\main{\alpha =-{81\over 10}.}
This is our main result in this paper.
With this exponent, it is obvious that the integral $\int d^9X
\rho (\X)$ is convergent at the origin, and divergent at
large $\X$. This is just fine, since our ansatz \sca\ is supposed
to be valid only for $|\X|< L$. We shall see shortly
that this restriction agrees with the black hole physics as governed by 
formula \free. Beyond the size $L$, we enter into the perturbative
region of quantum mechanics. For a threshold bound state, the
decay of the wave function at sufficiently large distance
also obeys a power law. The exponent in the asymptotic region
must be smaller than $-9$.

We still have to worry about the convergence of the integral in
\free. The function $g(x)$ damps fast for large $x$, peaks
at $x=0$. 
Integrating over $Y$ first, we shall obtain 
another factor $|\X|^\alpha $ for large $|\X|$, thus integration 
over $X$ is convergent for large $|\X|$. For small $\X$, integrating
over $\Y$ first will give rise to a order $O(1)$ number, so
integration over $\X$ near the origin is also convergent.
We conclude that the integral in \free\ is a finite
pure number.

The contribution to the integral in \free\ mainly comes from the region
$|\X|\sim L\lambda_{eff}^{-1/3}$, in terms of the coordinates before
rescaling. This size is smaller than $L$, but
independent of $N$. Thus it can not be
identified with the horizon radius \radius. There may be a few
different definitions of the horizon radius. One simple definition
is the subtracted quantity $1/N\langle\tr X^2\rangle$. By subtraction
we mean that the physical contribution to the average comes
only from the thermal fluctuations. This is reasonable, since
by scattering an object against a threshold bound state, the
wave function size should be not probed, as a basic assumption
in a holographic theory. Another more precise definition is
proposed in \kl. The proposal is that the horizon radius is where
a tachyonic mode develops between a probing D0-brane and the
black hole. The authors of \kl\ argue that the horizon radius is determined
by the first negative eigen-value of the following matrix
\eqn\mate{\langle M^2_{ij}\rangle =\langle \delta_{ij}
\X^2 +2[X_i,X_j]\rangle .}
It is expected that the negative eigen-value will be of order
$1/N\langle \tr X^2\rangle$. Since we are not trying to compute the
exact coefficient, we will not endeavor to use the above definition.

The amplitude $X^2$ of an off-diagonal element
stretched between point $\X$ and point
$\Y$ is $nl_p^3|\X-\Y|^{-1}$, when it is in its n-th harmonic
state. The thermal average is than
\eqn\themer{ {l_p^3\over |\X-\Y|}f(\beta R|\X-\Y|l_p^{-3}),}
where $f$ is the Boltzmann factor. We have the following thermal average
\eqn\tthem{{1\over N}\langle \tr X^2\rangle_\beta
={l_p^3\over N}\int d^9Xd^9Y\rho(\X)\rho(\Y)|\X-\Y|^{-1}f(\beta R|\X-\Y|
l_p^{-3}).}
Performing the same rescaling as before, we obtain
\eqn\otherm{{1\over N}\langle \tr X^2\rangle_\beta
={Nl_p^3\over L}\lambda_{eff}^{-4/15}\int d^9Xd^9Y
|\X|^{-81/10}|\Y|^{-81/10}|\X-\Y|^{-1}f(|\X-\Y|).}
Now, the integral is also convergent, since the superficial
pole in $|\X-\Y|^{-1}$ as well as in $f(|\X-\Y|)$ 
is not severe in 9 dimensions. Thus this integral is a pure number. 
Substituting
what we know about $L$ and $\lambda_{eff}$, the above quantity
is of order
\eqn\estim{N^{2/5}l_p^{18/5}(\beta R)^{-4/5}.}
This is precisely $r_0^2$, as in \radius. $r_0$ can be also
expressed as $L\lambda_{eff}^{-2/15}$. There are three scales appearing
in our problem
\eqn\thsca{L\gg L\lambda_{eff}^{-2/15}\gg L\lambda_{eff}^{-1/3}.}
We are fortunate that it is not the last scale being identified
with the horizon radius. The correct horizon radius we obtained
is a strong consistency check of our picture.

The next thing we wish to do is to identify the mass gap.
Our definition of the mass gap is the quantity $m^2$ appearing
in the following effective action
\eqn\effa{S={1\over R}\int dt\tr \left((\dot{X})^2
+m^2X^2+\dots \right),}
where we use the Euclidean time, the dots denote high order
terms. The quadratic term $\tr X^2$ comes
from thermal fluctuations. We expect the answer be $m^2\sim T^2$.
As explained in the appendix, this is going to be a 
nonperturbative result.

It is rather easy to see how to compute $m^2$ in our picture.
It is just the average mass square of stretched strings. From
the quadratic term $Rl_p^{-6}\tr [X_i,X_j]^4$ in the matrix action,
we find the right quantity to be averaged over is 
$R^2l_p^{-6}|\X-\Y|^2$. An additional factor $R$ comes from
our definition in \effa. Obviously, up to a numerical factor,
one can replace $|\X-\Y|^2$ by $(L\lambda_{eff}^{-1/3})^2$.
Therefore the mass gap is
\eqn\massg{m^2\sim R^2l_p^{-6}(L\lambda_{eff}^{-1/3})^2
=T^2.}
We see that all factors depending on $R$, $l_p$ and $N$
cancel. In a way, we should not be surprised by this
result.

It is possible to follow the strategy of \bfk\ to compute
the Hawking radiation rate. It may also be possible to use the exact
coefficient in the entropy formula to fix the coefficient in
the formula of D0-brane density, then coefficients in other
physical quantities may be computed within our picture.
We are only beginning to investigate the fascinating subject
of the large N D0-brane threshold bound state, and its role in
the duality between the large N D0-brane quantum mechanics
and M theory. We have only performed some qualitative analysis. 
Hard work remains to be done.

I understand that D. Kabat and G. Lifschytz have been investigating
the ten dimensional black holes along a different line.

\noindent {\bf Acknowledgments}
This work was supported by DOE grant DE-FG02-90ER-40560 and NSF grant
PHY 91-23780. Part of this work was carried out during a visit
at Rutgers University and a visit at KIAS at Seoul. Both institutions are
gratefully acknowledged for their warm hospitality. I have benefited
during the course of this work from conversations with T. Banks,
M. Douglas and P. Yi.

\noindent {\bf Appendix}

There are a few papers discussing D0-brane statistical mechanics
in the perturbative regime \stat. To our knowledge, the one-loop
mass gap has never been computed. We will give a formula at the
one-loop level here.

The Euclidean action of D0-brane quantum mechanics is
\eqn\eact{S_E=\int_0^\beta d\tau\tr\left( {1\over 2R}
(D_\tau\X)^2 -{R\over 16\pi^2l_p^6}[X_i,X_j]^2+\theta D_\tau\theta
+{R\over 2\pi l_p^3}\theta\gamma_i[X_i,\theta]\right).}
Performing rescaling $\tau\rightarrow \beta\tau$, $X_i
\rightarrow \sqrt{\beta R}X_i$, we reach the following action
\eqn\seact{S_E=\int_0^1d\tau\tr\left({1\over 2}(D_\tau\X)^2
-{g^2\over 4}[X_i,X_j]^2+\theta  D_\tau\theta
+g \theta\gamma_i[X_i,\theta]\right),}
where the effective coupling is $g^2=(\beta R)^3l_p^{-6}(4\pi^2)^{-1}$.

The mass gap is the coefficient appearing in the effective
quadratic potential. One way to compute the static potential is to use
the background field method. The first step is to fix the gauge.
In general, the gauge field $A$ can be put into a diagonal
form, each diagonal element represents a U(1) holonomy. We are
only interested in the mass gap, so we set $A=0$.
For fermions, since the action
is quadratic, one can even compute the exact static potential
coming from this part. The simplest way to compute it is to 
start with the Hamiltonian formalism. For a given static background
$X_i$, the Hamiltonian is simply
\eqn\fham{H_f={g\over 4}\tr \left(\theta\gamma_i[X_i,\theta]\right)
={1\over 2}\theta^a_\alpha {\cal N}_{a\alpha, b\beta}\theta^b_\beta,}
where $a,b$ are indices of the su(N) Lie algebra, and 
$\alpha, \beta$ are spinor indices. 
We have rescaled $\theta$'s such that $\{\theta^a_\alpha,\theta^b_\beta\}
=2\delta_{ab}\delta_{\alpha\beta}$.
The $16(N^2-1)\times 16(N^2-1)$
matrix ${\cal N}$ is
\eqn\fma{{\cal N}_{a\alpha, b\beta}={1\over 2}ig\gamma_i^{\alpha\beta}X_i^cf_{abc}.}
It is easy to check that ${\cal N}$ is an antisymmetric matrix.

Now $\theta^a_\alpha$ form a $16(N^2-1)$ dimensional Clifford algebra.
The fermionic Hilbert space is a spinor representation of this
algebra. The static potential is encoded in $\tr e^{-H_f}$. Since
${\cal N}$ is an even rank antisymmetric matrix, by an orthogonal rotation,
it can be put into the Jordan form. Once this done, we see that
\eqn\fcon{\tr e^{-H_f}=\left(\det (e^{\cal N}+e^{-{\cal N}})\right)^{1/2}.}
The determinant is invariant under the orthogonal rotation, thus the
above formula is valid for a general ${\cal N}$. For a small ${\cal N}$,
one can perform a perturbative expansion. The first term in
the effective action is quadratic in ${\cal N}$. Indeed we
have
$$\tr e^{-H_f}=e^{1/4\tr {\cal N}^2+\dots},$$
where the trace is taken over $(a,\alpha)$. In the end, we have
\eqn\ffac{\tr e^{-H_f}=e^{ Ng^2\tr\X^2+\dots} .}
Apparently, this expansion is valid when the coupling $\lambda_{eff}
=Ng^2$ is small. We see that the contribution of fermionic fluctuations
to the mass gap $m^2$ is negative.

It is a little more complicated to compute the contributions from
bosonic fluctuations. We separate bosonic matrices into a sum
of the background and the fluctuation $X_i+Y_i$. The one-loop
contribution is determined by the quadratic part in the fluctuation
$Y_i$ in the action
\eqn\bacc{S_b={1\over 2}(\p_\tau Y^a_i)^2 +{1\over 2}
Y^a_i{\cal M}_{ai,bj}Y^b_j,}
with 
\eqn\bma{{\cal M}_{ai,bj}=g^2[-i[X_i,X_j]^cf_{abc}
+X_k^cX^d_kf_{ace}f_{bde}\delta_{ij}-X_j^cX_i^df_{ace}f_{bde}].}
The path integral can be performed in two steps. First one converts
it into the form $\tr e^{-H_b}$. This gives the result
\eqn\nake{\tr e^{-H_b}= \left( \det[\sinh (\sqrt{{\cal M}}/2)]\right)^{-1}.}
This result is singular when ${\cal M}=0$. The problem is caused
by zero modes. We have already included the zero modes in the
background fields, the contribution of this part must be subtracted.
The subtracted result is then
\eqn\subt{{\det[\sqrt{{\cal M}}/2]\over \det [\sinh (\sqrt{{\cal M}}/2)]}.}
For small ${\cal M}$, the contribution to the effective action
from bosonic fluctuations is then
\eqn\bcon{-{1\over 24}\tr{\cal M}=-{1\over 3}Ng^2\tr \X^2+\dots .}
Although this contribution to the mass square is positive, its magnitude
is smaller than that from fermionic fluctuations. The overall
mass gap $m^2$ is negative. If we rescale back to the original time
and fields $\X$, we see that $m^2\sim Ng^2T^2\sim \lambda_{eff}T^2$.

It is not surprising to have a negative mass square at a finite 
temperature. The presence of temperature breaks supersymmetry, thus
the cancellation between bosonic fluctuations and fermionic
fluctuations is spoiled. The phase space of bosons is overwhelmed by 
that of fermions, that at short distance there is a repulsive
force. This agrees with the result for two D0-branes in the last
paper of \stat, where integration over the holonomy of the gauge
field is also taken into account. We expect that high order
terms in the static potential takes over  for large $\X$, thus
the net force will be attractive.

As we have seen in the main body of this paper, the mass gap
$m^2$ is positive in the strong coupling regime, and is independent
of $\lambda_{eff}$. When one tunes down $\lambda_{eff}$, there
must be a point where the mass gap vanishes. One might take this
point as a phase transition point. This phase transition is
generally predicted in \gl.

\vfill
\eject

\listrefs
\end